\documentclass[fleqn,usenatbib,letters]{mnras}

\usepackage{newtxtext,newtxmath}

\usepackage[T1]{fontenc}

\DeclareRobustCommand{\VAN}[3]{#2}
\let\VANthebibliography\thebibliography
\def\thebibliography{\DeclareRobustCommand{\VAN}[3]{##3}\VANthebibliography}

\usepackage{graphicx}	
\usepackage{amsmath}	

\newcommand{\OO}{\={O}tautahi--Oxford model}
\newcommand{\psj}{PSJ}
\usepackage{hyperref,xurl} 

\title[3I/ATLAS is red]{Snapshot of a new interstellar comet: 3I/ATLAS has a red and featureless spectrum.}

\author[C. Opitom et al.]{Cyrielle Opitom,$^{1}$\thanks{E-mail: copi@roe.ac.uk (CO)}
Colin Snodgrass,$^{1}$
Emmanuel Jehin,$^{2}$
Michele T. Bannister,$^{3}$
Erica Bufanda,$^{1}$
\newauthor
Sophie E. Deam,$^{4,5}$
Rosemary C. Dorsey,$^{3,6}$ 
Marin Ferrais,$^{7}$
Said Hmiddouch,$^{2,8}$
Matthew M. Knight,$^{9}$
\newauthor
Rosita Kokotanekova,$^{10,11}$
Brayden Leicester,$^{3}$
Micha\"el Marsset,$^{12}$
Brian Murphy,$^{1}$
Vincent Okoth,$^{1}$
\newauthor
Ryan Ridden-Harper,$^{3}$
Mathieu Vander Donckt,$^{2}$
Léa Ferellec,$^{1}$
Damien Hutsemekers,$^{2}$
Manuela Lippi,$^{13}$ 
\newauthor
Jean Manfroid,$^{2}$
Zouhair Benkhaldoun,$^{8}$
\\
$^{1}$Institute for Astronomy, University of Edinburgh, Royal Observatory, Edinburgh EH9 3HJ, UK\\
$^2$ STAR Institute, University of Liège, Allée du 6 août, 19, 4000 Liège (Sart-Tilman), Belgium\\
$^3$ School of Physical and Chemical Sciences -- Te Kura Mat\={u}, University of Canterbury, Private Bag 4800, Christchurch 8140, New Zealand\\
$^{4}$Space Science and Technology Centre, School of Earth and Planetary Sciences, Curtin University, Perth, Western Australia 6845, Australia \\
$^{5}$ International Centre for Radio Astronomy Research, Curtin University, Perth WA 6845, Australia \\
$^6$ Department of Physics, University of Helsinki, P.O. Box 64, 00014
Helsinki, Finland\\
$^{7}$ Florida Space Institute, University of Central Florida, 12354 Research Parkway, Orlando, FL 32828, USA \\
$^8$ Cadi Ayyad University (UCA), Oukaimeden Observatory (OUCA), Faculté des Sciences Semlalia (FSSM), Marrakech, Morocco\\
$^9$ Physics Department, United States Naval Academy, 572C Holloway Road, Annapolis, MD 21402, USA\\
$^{10}$ Institute of Astronomy and National Astronomical Observatory, Bulgarian Academy of Sciences, 72 Tsarigradsko Shose Boulevard, 1784 Sofia, Bulgaria\\
$^{11}$ International Space Science Institute, Hallerstrasse 6, 3012 Bern, Switzerland\\
$^{12}$ European Southern Observatory (ESO), Alonso de C\'ordova 3107, 1900 Casilla Vitacura, Santiago, Chile \\
$^{13}$ INAF, Osservatorio Astrofisico di Arcetri, Largo E. Fermi 5, 50125, Firenze, Italy
}

\date{Accepted XXX. Received YYY; in original form ZZZ}

\pubyear{\the\year{}}

\begin{document}
\label{firstpage}
\pagerange{\pageref{firstpage}--\pageref{lastpage}}
\maketitle

\begin{abstract}
The interstellar comet 3I/ATLAS is only the third interstellar object to be discovered. 
Pre-perihelion measurements provide a unique opportunity to study its activity and composition, which may alter as it is heated in the coming months. 
We provide an initial baseline from optical spectroscopic observations obtained only two days after discovery, using the MUSE instrument on the VLT on 2025 July 3, while 3I was at 4.47 au from the Sun and 3.46 au from the Earth. 
These observations confirm the cometary nature of 3I, and reveal a red coma with a spectral slope of $(18\pm4)\%/1000$~\AA, redder than most Solar System comets but similar to the surface colour of some Solar System Trans-Neptunian Objects or Centaurs. 
We searched for but did not detect gas emission from C$_2$, NH$_2$, CN, and [OI], which is consistent with volatile non-detections for Solar System comets at this heliocentric distance.
At present, the coma appears entirely dusty.
Future observations of 3I as it comes closer to the Sun will provide an invaluable opportunity to witness the evolution of its activity, study its composition, test predictions of interstellar object population models, and compare 3I to Solar System comets. 
\end{abstract}




\section{Introduction}

The study of macroscopic interstellar objects (ISOs) in our Solar System began with the discovery of 1I/`Oumuamua in 2017 \citep{Meech2017}, followed by 2I/Borisov in 2019 (1I and 2I, respectively, hereafter). 
These small bodies were remarkably different, the former showing no obvious activity while the latter had a clear coma. 
1I demonstrated a number of surprising features, such as a large lightcurve amplitude \citep{Knight:2017, Bannister:2017, Fraser:2018} and comet-like non-gravitational acceleration without a detectable coma (\citealt{Micheli:2018}; see \citealt{ISSITeam:2019} for a review). 
2I instead appeared quite similar to Solar System comets, showing typical cometary spectral features \citep{Fitzsimmons:2019,Opitom:2019,McKay:2020,Kareta:2020,Bodewits:2020, Bannister:2020, Opitom:2021,Deam:submitted}.

The discovery of a third ISO has been eagerly awaited, in part to understand the diversity of ISOs throughout the Galaxy. 
ISO population models predict a chemodynamic range of physical properties that relate to host star metallicity \citep{Lintott_2022, Cabral_2023, Hopkins2023}. 
In the \OO, for ISOs formed in the disc outside their systems' water ice line, water mass fraction should show a correlation with stellar metallicity \citep{Hopkins2025}.  
A comparatively small sample of this large galactic population pass through the Solar System. 
The Vera C. Rubin Observatory is expected to find a significant number of ISOs (6-51 over 10 years of surveying) due to its greatly increased sensitivity compared with current sky surveys \citep{Dorsey:2025}, but it is just beginning its commissioning operations.  
In the interim, a third ISO has been discovered by the 0.5m-aperture ATLAS survey \citep{Tonry2011}.

3I/ATLAS (hereafter 3I), like 2I, is cometary. 
It was quickly found to have low-level coma, and also designated C/2025 N1 (ATLAS), following initial reports of activity\footnote{\url{https://minorplanetcenter.org/mpec/K25/K25N12.html}}, supported by independent reports from \citet{Jewitt-ATEL,ATel_17264,Seligman:2025}. 
Discovered at just under 5~au from the Sun, 3I has a substantial velocity at infinity of close to 60~km/s \citep{Hopkins:submitted}, and will reach a perihelion of 1.35~au in late October 2025. 
It will be visible to Earth-based observers for around a year, with a $\sim$2 month gap in observability around its perihelion when it will have Solar elongation below 50$^\circ$. 
3I presents an excellent opportunity to study the evolution of an ISO as it approaches and crosses the region around the Sun where equilibrium temperatures are high enough for significant water ice sublimation (typically, within $\sim$3 au; e.g. \citealt{Meech:2004}), and over a range of distances where Solar System comets have historically been observed.
This will allow direct comparisons with our own Solar System's native population. 

In this Letter, we present initial characterisation observations of 3I/ATLAS obtained with the ESO VLT and the MUSE instrument \citep{Arsenault2008}. 
MUSE is an integral-field-unit spectrograph covering the wavelength range 4800-9200 \AA, allowing us to obtain both spectroscopy and 2-D maps of any detected coma species. 
MUSE, in the Wide Field Mode used for these observations, has a field of view of 1\arcmin$\times$1\arcmin$ $. 
It has previously been demonstrated to be a powerful tool for characterising comets and active asteroids \citep[e.g. ][]{Opitom:2020,Opitom:2023,Kwon:2023,Murphy:2023}, and was successfully used to observe 2I for 16 epochs through perihelion \citep{Deam:submitted}.

\section{Observations}

Observations of 3I were triggered in target-of-opportunity mode the night after the announcement of the discovery of the ISO, taking place on 2025 July 3rd between 00h25 and 01h12 UT. 
At this time 3I was at 4.47 au from the Sun, 3.46 au from Earth, and at a phase angle of 2.5$^\circ$. 
3I was crossing a crowded star field, with a galactic latitude of only 1.6$^\circ$. 
To minimise contamination by the dense field stars, we used an exposure time of 300 seconds and took 8 exposures over one hour, with small dithers and $90^\circ$ rotation between exposures.
This allowed us to correct detector-to-detector effects and combine the spectra of the target to remove background stars.

We reduced the data using the standard MUSE pipeline \citep{Weilbacher2020}, including sky subtraction, telluric correction, and flux calibration using the spectrophotometric standard observed during the same night. 
We performed a dedicated search for [OI] emission at 6300~\AA, using the methodology described in \citet{Opitom:2020}. 
Despite the unusually high Doppler shift relative to Earth for a comet at this heliocentric distance (owing to its interstellar excess velocity), the cometary oxygen lines were still not fully resolved from the telluric lines. 
We took advantage of the relatively large field of view of MUSE to create maps of [OI] where cometary emission should manifest as an enhancement around the centre of the field, contrasting with the background pattern. 

We extracted spectra using a $1\arcsec$ radius aperture, to avoid contamination from background stars, and median combined seven of the eight individual spectra. 
The sixth spectrum was discarded due to strong contamination from a background star. 
Regions strongly affected by telluric absorption that could not be completely corrected by the pipeline were dismissed from the analysis. 
A reference solar spectrum from the SOLar SPECtrmeter (SOLSPEC) instrument of the SOLAR payload on board the International Space Station \citep{Meftah2018} was used to perform dust continuum subtraction and compute the spectrum reflectance. 

In addition, photometric observations with $B$, $V$, $Rc$, and $Ic$ filters were obtained on the nights of July 3rd to July 6th with the TRAPPIST-North telescope (TN; Z53) located at the Ouka{\"i}meden observatory in Morocco. 
TN is a 0.6-m Ritchey-Chr{\'e}tien telescope operating at $f/8$. 
It is equipped with an Andor IKONL BEX2 DD camera providing a 20$\arcmin$ field of view and pixel scale of 0.60$\arcsec$/pixel. 
We acquired 5 images of 180s per filter and we used a 2$\times$2 binning (1.2$\arcsec$/pixel).  
Each image was carefully examined, and images where the target was contaminated by background stars were discarded. 
The photometry was performed with the Photometry Pipeline \citep{Mommert:2017} by matching in each image typically 100 field stars with the PanSTARRS DR1 photometric catalogue. 
An aperture of 4 pixels radius was used.

\section{Results}

\begin{figure}
\centering
	\includegraphics[width=0.8\columnwidth]{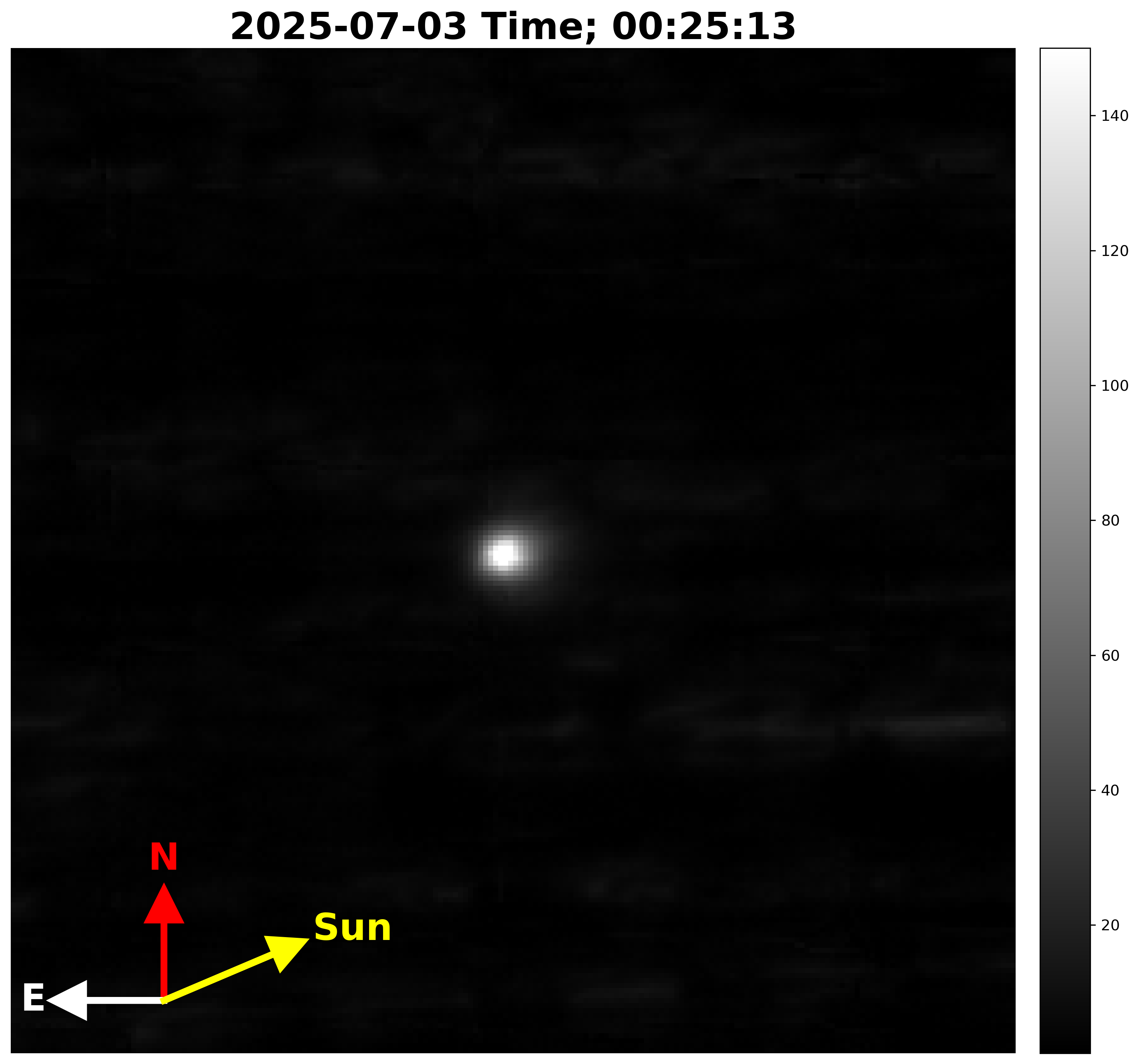}
    \caption{Combined white light image of 3I/ATLAS, created by median stacking white light images produced by collapsing in wavelength individual data cubes. The field of view displayed is $40\arcsec \times 40\arcsec$. Intensities are in units of 10$^{-20}$ erg/s/cm$^2$.}
    \label{fig:image}
\end{figure}

The spatial and wavelength ($x,y,\lambda$) data cubes produced by the MUSE pipeline were each collapsed in wavelength to produce white light images, followed by taking the median of all eight images to produce a deep stacked image of 3I and minimise the impact of background stars (Fig.~\ref{fig:image}). 
This shows a slightly extended object (with a full width at half maximum $\sim2$\arcsec, compared to reported seeing at zenith of $\sim1$\arcsec{} at the time of the observations), extended towards the West (position angle $\sim 290^\circ$). 
Our data confirm previous reports of cometary activity in 3I; these VLT images are the largest-aperture data reported to date, and provide a clear detection of a coma. 
The low phase  and  orbit plane (0.9$^\circ$) angles of the observations and relatively large distance means that any tail will be foreshortened and behind the comet in this geometry.
It is likely to become more obvious as the geometry changes and activity levels increase as the comet approaches the Sun.

\begin{figure}
\centering
	\includegraphics[width=0.9\columnwidth]{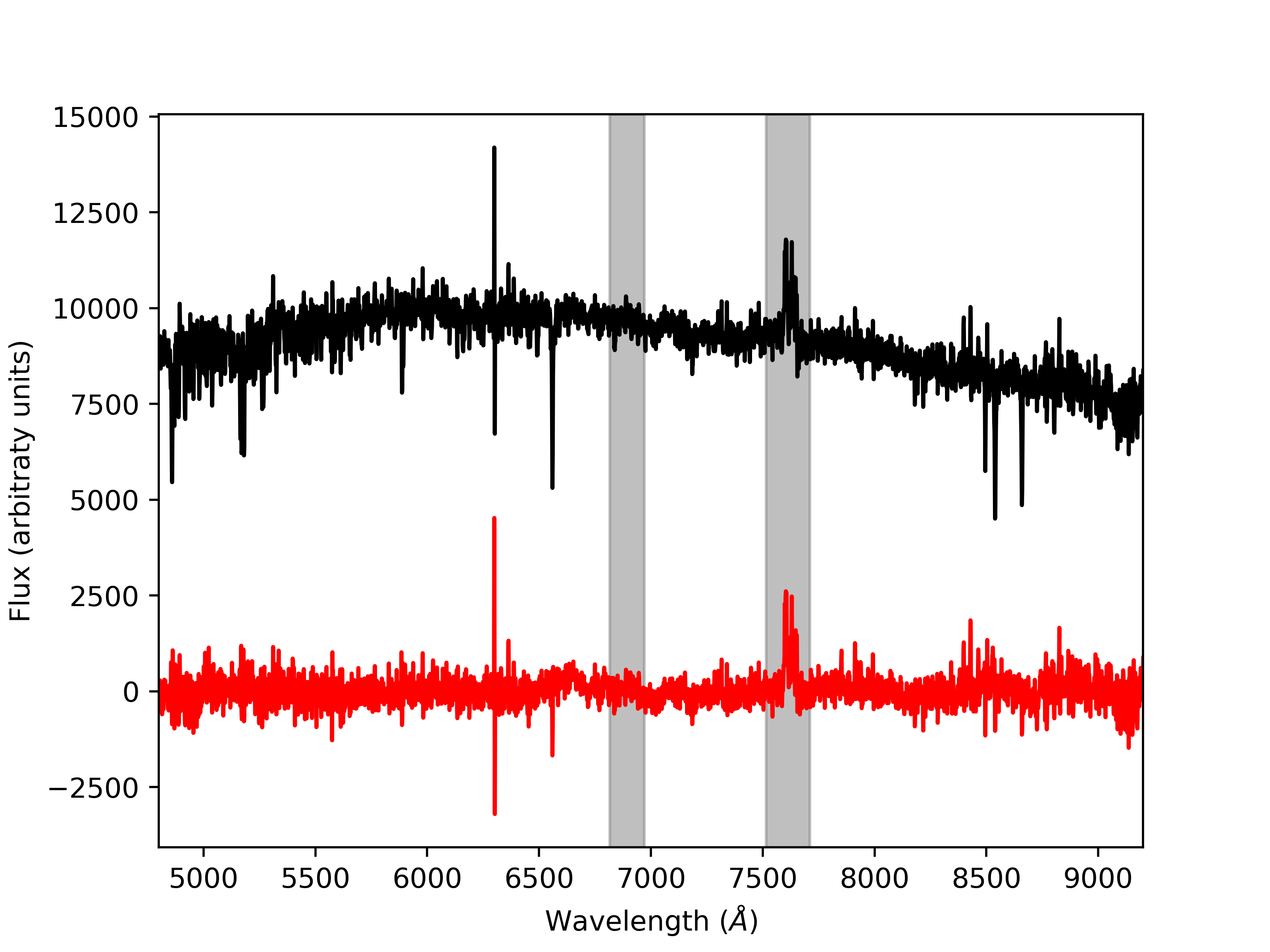}
    \caption{Black: Extracted spectrum in a 1\arcsec radius aperture. The spectrum is a median combination of the 7 uncontaminated spectra extracted from the reduced datacubes. Red: Continuum-subtracted spectrum. The vertical grey bands denote regions of the spectrum affected by telluric feature.}
    \label{fig:spectrum}
\end{figure}

The extracted spectrum is shown in  Fig.~\ref{fig:spectrum}. 
It shows the expected solar continuum, without any clear trace of gas emission. 
The wavelength coverage of MUSE includes emission from C$_2$ (the brightest being $(\Delta v=0)$), various bands of NH$_2$, and the red CN(1-0) band. 
None of these emissions are detected in this early spectrum of 3I. 
A specific search for [OI] emission, built to robustly separate sky and comet contributions in dedicated maps, did not reveal any gas emission.

\begin{figure}
\centering
	\includegraphics[width=0.8\columnwidth]{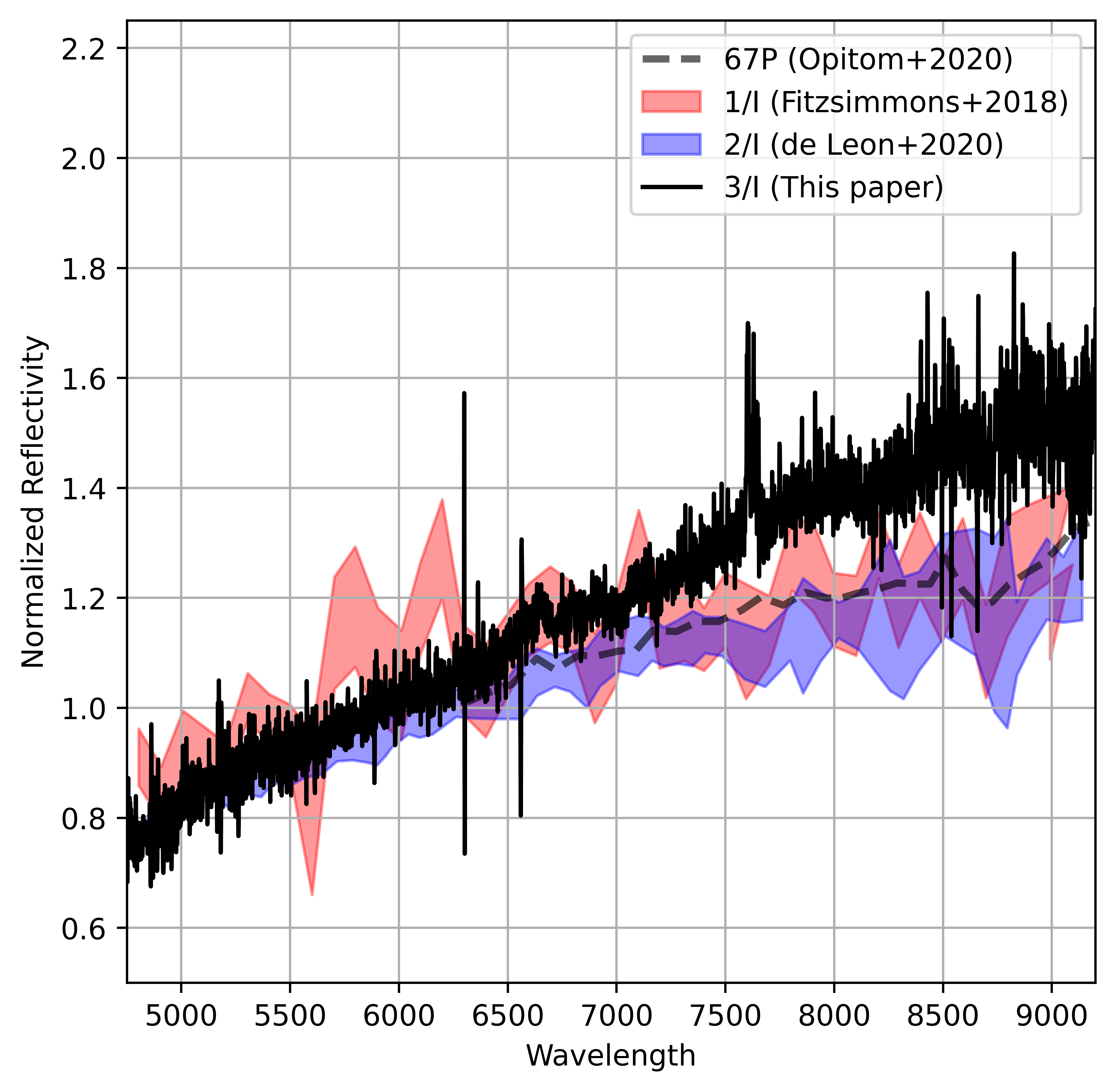}
    \caption{Optical reflectance spectrum of 3I compared to 1I \citep{Fitzsimmons:2019} and 2I \citep{de-Leon:2020} and 67P \citep{Opitom2020}.}
    \label{fig:reflectance}
\end{figure}

The reflectance spectrum of 3I was computed by dividing the median comet spectrum by a solar spectrum from \citet{Meftah2018} and normalising at 6000~\AA. 
It is displayed in Fig. \ref{fig:reflectance}, where it is also compared to those of the first two interstellar objects, 1I and 2I. 
It indicates a reddening through the entire wavelength range, without any significant features. 
Fitting a slope to the reflectance spectrum, we measure a normalised reflectivity gradient of $(18\pm3)\%/1000$~\AA\ in the 5000-7000~\AA\ range, $(17\pm4)\%/1000$~\AA\ in the 7000-9000~\AA\ range, and $(18\pm4)\%/1000$~\AA\ in the 5000-9000~\AA\ range. 
Colours measured from TRAPPIST broadband photometry are $B-V = 1.12 \pm 0.14$ , $V-R = 0.57 \pm 0.09$, $V-I = 1.10 \pm 0.15$, which correspond to a mean reflectivity gradient of $15\%/1000$~\AA\ in the 5400-7865~\AA\ range, with the spectroscopy results.

\section{Discussion and Conclusions}

\begin{figure*}
	\includegraphics[width=\linewidth]{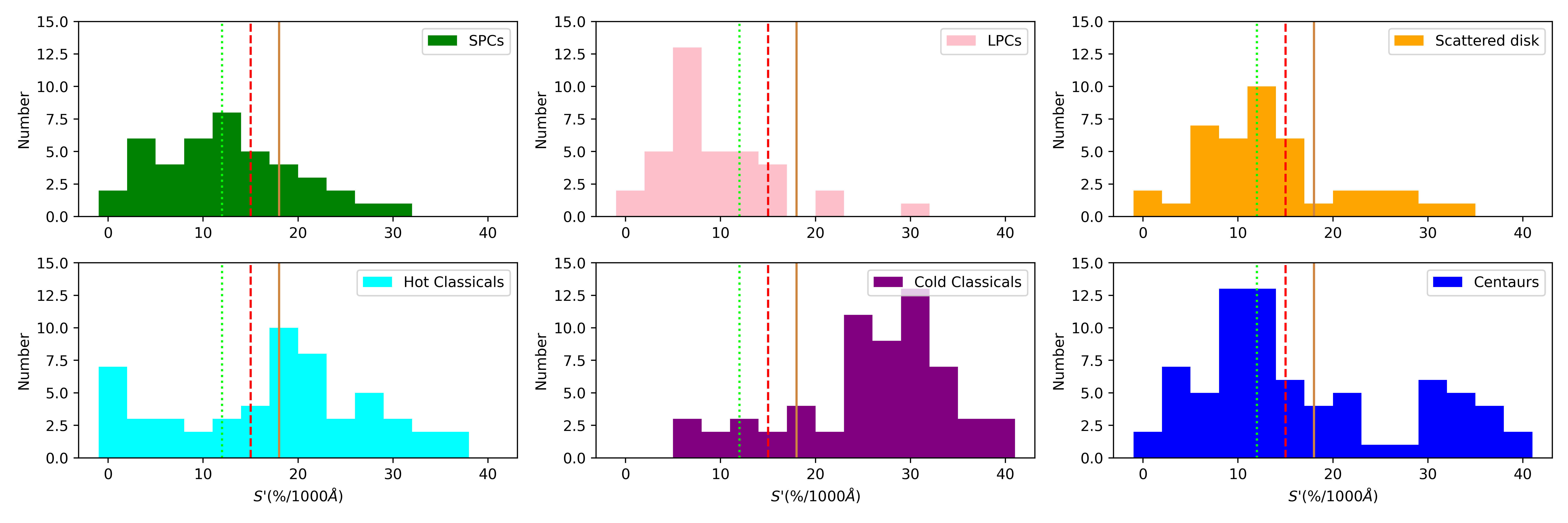}
    \caption{Normalised spectral slopes (in $\%/1000$\ \AA) for the three ISOs compared with various Solar System populations. Based on a figure by \citet{Jewitt:2023}, using data from \citet{Hainaut:2012}. Vertical dashed, dotted, and full lines represent the values for 1I, 2I (both taken from the review by \citealt{Jewitt:2023}), and 3I (this work) respectively.}
    \label{fig:specslopes}
\end{figure*}
 
3I exhibits a red colour.
The spectral reflectance is likely dominated by its dust coma, given that the object has been found to be active. 
In Fig. \ref{fig:reflectance} and Fig. \ref{fig:specslopes}, we compare the colour of 3I to that of the other two known interstellar objects and to various other Solar System populations. 
The colour and normalised reflectivity reported for 2I varied widely (5-22$\%/1000$~\AA), depending on the technique, epoch and wavelength range (\citealt{Fitzsimmons:2019,Guzik:2020,Yang:2019-updatedbyYang:2020,Kareta:2020,Mazzotta-Epifani:2021,Aravind:2021,Prodan:2024,Hui:2020,de-Leon:2020,Lin:2020}; see summary figure in \citealt{Deam:submitted}). 
The review by \citet{Jewitt:2023} adopts a nominal value of $(12\pm1)\%/1000$~\AA.
The reddest values were measured for 2I in the 3900-6000~\AA\ range \citep{Fitzsimmons:2019,deLeon2023}. 
Spectroscopic measurements from \citet{Kareta:2020,Prodan:2024,Lin:2020} over a similar wavelength range to the one covered here (5260-7130~\AA, 5700-7250~\AA, and 5500-9000~\AA\ respectively), yield generally bluer values than that measured for 3I in this work. 
In comparison, the reflectivity of 1I ranges from 7 to 23 $\%/1000$~\AA\ \citep{Jewitt:2017,Meech2017,Ye:2017,Fitzsimmons:2018}. 
\citet{Jewitt:2023}'s review chooses a value of $(15\pm5)\%/1000$~\AA\ for 1I. In the case of 1I, this reflects the colour of the directly-observable nucleus rather than any dust coma, while the colour measurements for 2I and 3I likely represent the dust coma.
\citet{Seligman:2025} also report $g'r'i'z'$ colours with a slope of $\sim 18\%/1000$~\AA$ $ and spectroscopy with a slope of $(17.1\pm0.2)\%/1000$~\AA\ for 3I, consistent with our results.

The colour of 3I appears redder than what is usually measured for the coma of most Solar System comets, and potentially closer to that reported for the surfaces of some TNOs or Centaurs, as illustrated in Figure \ref{fig:specslopes}. 3I sits somewhere in the middle of the distribution of dynamically excited TNOs and Centaurs; it is redder than most comets but bluer than the very red objects in the cold classical Kuiper Belt. One plausible interpretation of 3I having a  redder colour than most comets or active Centaurs is that we are seeing the beginning of its activity, and the coma represents surface layers similar to the surfaces of outer Solar System bodies, rather than fresher material from the interior. 
It has been suggested that activity causes a loss of ultrared matter, explaining the difference between comets and TNOs \citep{Jewitt:2002,Snodgrass:2006}. Supporting this idea, the transition Centaur population is described by a colour bimodality, where the active Centaurs belong to the less red colour type \citep[e.g.][]{Peixinho2025,Kokotanekova2025}. An intriguing exception is 523676 (2013 UL10), active at 6.2 au, and which is the first active Centaur to display a red colour, contrasting with the generally active Centaurs \citep{Mazzotta-Epifani:2018}. 3I is redder than the average of active Centaurs but bluer than 2013 UL10. 
On their journey, interstellar objects will be subjected to radiation bombardment, which will alter their surfaces. 
The interstellar radiation environment should be similar to that experienced by Solar System comets in the Oort cloud, which is well beyond the heliopause \citep{Stern:2003}, so one might expect similarities for ISOs to long period comets (LPCs) rather than short period comets (SPCs) with a source region in the TNOs. 
However, 3I appears to be redder than most LPCs (Fig.~\ref{fig:specslopes}), and in any case the differences between the average colours of LPCs and SPCs are small \citep{Jewitt:2015}.

It is worth noting that the MUSE red spectral slope of 3I is in agreement with the spectral slope presented by \citet{Seligman:2025}, and our average TRAPPIST colours. However, some of the available TRAPPIST photometric colours vary from very red at shorter wavelengths to more neutral at longer wavelengths, and the colours reported by \cite{Bolin2025-arxiv} point to a  more neutral slope. We attribute these significant differences to the challenging nature of the observations during the first days since the discovery of 3I. Crossing the galactic plane, the object has been captured in crowded fields where stellar contamination has to be carefully accounted for in the photometric analysis. In these fields, extra care needs to be taken with the colour correction performed during the absolute calibration of the frames since extinction in the galactic plane is known to affect the colour distribution of comparison stars in the frame. Another factor that might contribute to the noticeable differences in the spectral slope and the photometric colour indices is the increasing proximity to the moon which might lead to possible contamination in the bluer filters. For these reason, we consider the spectroscopic slope more reliable.


Based on previous reports and our data, 3I is clearly active, with a compact coma extended approximately in the sunward direction (position angle $\sim 290^\circ$). 
Within the limits of the current data (MUSE is not an imager and at $m_r \sim 17$, 3I is comparatively faint), it is difficult to confirm whether the extension fully coincides with the Sun's direction. 
However, this approximately sunward feature is reminiscent of the distant activity of other comets (e.g. C/2014~UN271, \citealt{Farnham:2021}) that had activity dominated by large particles emitted in the sunward direction.

The non-detection of cometary emission is reasonable, given 3I's large heliocentric distance at the time of the observations. 
Although gas has been occasionally detected at comparable or even larger distances from ground-based optical observations (cf. detections of CN in Centaur 29P/Schwassmann-Wachmann~1 by \citealt{Cochran:1991} at 5.8~au; post-perihelion detections in Oort cloud comet C/1995 O1 Hale-Bopp of C$_3$ at 7.0~au and CN at 9.8~au by \citealt{Rauer:2003}; pre-perihelion detections in Oort cloud comet C/2001 Q4 NEAT of [OI] at 3.7~au by \citealt{Decock:2013}), these detections were for comets considerably brighter (i.e. 4--5 magnitudes) than 3I. 
Our target-of-opportunity observations would likely have needed to be substantially longer to set meaningful constraints on gas activity. 

\citet{Hopkins:submitted} predict a high water mass fraction for 3I, based on its chemodynamics: from the \OO, 3I is the first ISO from the Milky Way's thick disk population yet seen, and its origin star's metallicity would imply a higher water mass fraction than that of other ISOs.
Future MUSE epochs will test if e.g. [OI] becomes apparent as 3I approaches the Sun.
It is worth noting that the high incoming velocity of 3I relative to the Sun means that any thermal lag could cause a delayed onset of activity, relative to what might be observed in Solar System comets at similar heliocentric distances. 
Detailed thermal modelling, similar to that applied to 1I by \citealt{Fitzsimmons:2018}, will be necessary.


Recent work by \citet{Holt:2024} and \citet{Lacerda:2025} compiled brightness curves of Oort cloud comets beginning at distances comparable to or beyond the distance at which 3I was discovered. 
Consistent with earlier work \citep[e.g.,][]{Whipple:1978} that was restricted to smaller heliocentric distances, both found that dynamically new comets brighten more slowly than comets that had previously been through the inner Solar System. 
However, both found that the rate of brightening for all comets decreased as comets got closer to the Sun. 
3I's large discovery distance and relatively low perihelion distance mean that it will be intensely studied on its inbound track over a much wider range of thermal conditions than was 2I (solar radiation increases from discovery to perihelion by a factor of $\sim$11 for 3I versus only $\sim$2.3 for 2I, and 2I had a higher perihelia than will 3I), allowing for a comparison with solar system comets brightening rates.
However, its perihelion is great enough that 3I is very unlikely to disintegrate, though it may experience outburst events, as did 2I \citep{Drahus:2020,Jewitt:2020}.

3I presents the best opportunity yet to compare the evolution of an interstellar comet with those native to the Solar System --- potentially providing further insight into whether its activity is driven by the same mechanisms, and possibly even revealing if it has spent considerable time sublimating near another star \citep{Gkotsinas2024}, which 2I did not \citep{Bodewits:2020, Cordiner:2020, Deam:submitted}. 
Our ongoing VLT/MUSE programme will monitor the evolution of its spectral slope and gas composition, thus providing key data points for placing 3I into context with Solar System comets.



\section*{Acknowledgements}

Based on observations collected at the European Southern Observatory under ESO programmes 115.27ZJ.001. 
The authors wish to acknowledge the exceptional level of support provided by support astronomers and Instrument and Telescope Operators, in particular Susana Cerda-Hernandez and Thomas Szeifert at Paranal Observatory. 

For the purpose of open access, the authors have applied a Creative Commons Attribution (CC BY) licence to any Author Accepted Manuscript version arising from this submission. 

CO acknowledges the support of the Royal Society under grant URF\textbackslash R1\textbackslash211429.
M.T.B. appreciates support by the Rutherford Discovery Fellowships from New Zealand Government funding, administered by the Royal Society Te Ap\={a}rangi. 
R.R.-H. is supported by the Rutherford Postdoctorial Research Fellowship, and a Marsden Fast Start, administered by the the Royal Society Te Ap\={a}rangi.

The views expressed in this article are those of the authors and do not reflect the official policy or position of the U.S. Naval Academy, Department of the Navy, the Department of Defense, or the U.S. Government.
\section*{Data Availability}

Raw data will be available via the ESO archive at \url{https://archive.eso.org/wdb/wdb/eso/eso_archive_main/query?prog_id=115.27ZJ.001} following a proprietary period.



\bibliographystyle{mnras}
\bibliography{3I} 








\bsp	
\label{lastpage}
\end{document}